\documentstyle[prl,preprint,
aps,graphicx]{revtex}
\begin{document}
\draft
\preprint{ETH-TH/98-22}
\title{Primordial Black Holes from the QCD Transition?}
\author{Peter Widerin and Christoph Schmid \thanks{e-mail:
        widerin@itp.phys.ethz.ch, chschmid@itp.phys.ethz.ch}} 
\address{Institut f\"ur Theoretische Physik, 
         ETH-H\"onggerberg, CH-8093 Z\"urich}
\date{\today}
\maketitle
\tighten

\begin{abstract}
Can a violent process like sudden reheating after supercooling at the onset of
a first-order QCD transition improve the possibility of primordial black hole
formation? Underdensities reheat earlier than overdensities, there is a short
period of huge pressure differences, hence fluid acceleration. Density
perturbations on scales far below the Hubble radius $\lambda\ll R_{\rm H} $
get an amplification which grows quadratically in wavenumber, the
amplifications at the horizon scale are small. Primordial black hole formation
cannot be sufficiently amplified by the QCD transition unless the initial
spectrum is fine tuned.    
\end{abstract}
\narrowtext

\section{Introduction and Conclusion}

The transition from a quark-gluon plasma to a hadron gas in the early universe
took place at a temperature $T_{\star} \sim 150$ MeV. Recent results of
lattice QCD indicate a first-order phase transition for the physical values of
the $u$, $d$, $s$ quark masses \cite{lattice}. The mass inside the Hubble
horizon at the QCD transition is $\sim 1 M_{\odot}$. Crawford and Schramm
\cite{Schramm} first proposed that black hole formation at the QCD transition
could account for dark matter today. Recently Jedamzik \cite{Jedamzik}
proposed to identify such primordial black holes (PBH) with the MACHO's
(massive compact halo objects) observed by microlensing
\cite{microlensing}. He pointed out that the formation of PBH's should be
particularly efficient during the QCD transition due to a significant decrease
in the effective sound speed during the transition. Schmid, Schwarz and
Widerin \cite{Helv,PRL} showed that during a first-order QCD transition the
sound velocity $c_s=\left(\partial p/\partial \rho\right)_s^{1/2}$ vanishes
(for wavelengths much larger than the bubble separation), there are no pressure
gradients, no restoring forces, and preexisting cosmological density
perturbations go into free fall. They computed the amplification of linear
cosmological perturbations due to the QCD transition. The amplification is
very large for scales $\lambda$ far below the horizon, amplification $\propto
k^n$ with $n=1$ in the bag model, $n=3/4$ in a fit to lattice QCD data. There
is no amplification for superhorizon modes in accordance with a general
theorem. Near the horizon the first peak (above the incoming spectrum of
cosmological perturbations) has a height (amplification factor) of only 1.5
for the fit to the lattice QCD data and 2.0 for the bag model. The position of
the first peak is at $M_{\rm RAD}\approx 0.15 M_{\odot}$, which corresponds to
$\lambda/R_{\rm H} \approx 0.6$ at the end of the first order QCD
transition. It was concluded that primordial black hole formation is unlikely, 
because the amplifications of perturbations due to the QCD transition are
large only far below the Hubble scale.
 
In this paper we ask, whether a violent process like sudden reheating after
supercooling at the beginning of the QCD transition could enhance the
amplification of preexisting density perturbations and improve the possibility
of black hole formation at the QCD transition.

The amount of supercooling $(T_\star - T_{\rm sc})/T_\star \equiv \eta$ is
proportional to the dimensionless ratio $(\sigma^{3/2}/(l T_\star^{1/2})$,
where $\sigma =$ surface tension, $l =$ latent heat, see e.g. \cite{PRD}. Data
are available only for quenched lattice QCD (gluons only, no quarks)
\cite{Iwasaki}: $l/T_\star^4 \approx 1.40 (9)$ and a very small surface tension
$\sigma/T_\star^3 \approx 0.0155 (16)$. This gives a very small supercooling
of $\eta \approx 10 ^{-3}$ \cite{PRD}. In view of the uncertainties in the
lattice QCD determinations and the absence of data for $l$ and $\sigma$ from
QCD including three physical quarks we consider the possibility of a larger
surface tension and/or a smaller latent heat giving more supercooling. For
illustrative purposes in our figures we used $\eta = 1/10$. 

The amplification of perturbations due to sudden reheating at the beginning of
a first order QCD transition occurs because reheating happens at a certain
supercooling temperature $T_{\rm sc}$. Hence underdensities reheat earlier
than overdensities. (For subhorizon scales Newtonian time is applicable.)
During this short period of time, there are spatial variations of pressure of
${\mathcal O}(\eta)$ leading to huge pressure gradients (compared to the the
small preexisting cosmological pressure gradients) and huge fluid
accelerations, hence preexisting overdensities get a very violent but very
short compressional impulse. The fluid velocity effectively jumps proportional
to the wave number $k$. For scales far below the Hubble radius this jump
becomes the dominant contribution to the fluid velocity throughout the
transition. In the bag model \cite{bag}, which we shall consider here for
simplicity, the resulting amplification of density perturbations for scales
far below the horizon grows as $k^2$, i.e. with one additional enhancement
factor $\propto k$ due to sudden reheating. For horizon and superhorizon
scales general relativity must be used.  
 
The general relativistic time-evolution requires the choice of a foliation of
space-time, and for linear perturbations this defines a gauge. The
hypersurface of sudden reheating is a slice of constant energy density, and we
use the corresponding gauge, the uniform density  (UD) gauge, to evolve
through sudden reheating. In this gauge the relevant variables are continuous
at sudden reheating. To evolve before and after reheating the UD gauge is not
suitable, because it is singular in the subhorizon limit, instead we use the
uniform expansion (UE) gauge, which is nonsingular, and which is defined by
requiring that the fundamental observers, who are at rest on a slice, have a
uniform Hubble expansion. In the UE gauge the relevant variables are
discontinuous at sudden reheating, and these discontinuities are obtained in a
very simple way by using the gauge transformation between the UD and the UE
gauges. In the superhorizon limit the state variables in UE gauge do not
jump. Moreover we prove the theorem that phase transitions do not affect at
all the laws of evolution for $\delta \equiv \delta \rho/\rho$ of the growing
mode (i.e. growing in relative importance) in UE gauge in the limit $k^2_{\rm
phys} \ll \{H^2,\dot{H}\}$, 
\begin{equation}
\label{theorem}
\left(\delta_{\rm UE}\right)_{\rm growing \; mode} = {\rm const} \left({k_{\rm phys} \over
H}\right)^2.
\end{equation}
This translates into the 'conservation law' $\left(\varphi_{\rm
UE}\right)_{\rm growing \;mode} = {\rm const}$, where $4 \bigtriangleup
\varphi_{\rm UE} \equiv - ^{(3)}R(\Sigma_{\rm UE})$, by using the energy
constraint. Because of this theorem the superhorizon modes during the QCD
transition remain unaffected in the sense of Eq.~(\ref{theorem}). Near the
horizon the first amplification peak gives a factor $2.2$ for a supercooling
of $10 \%$ in the bag model, the additional enhancement due to sudden
reheating at the horizon scale is about $10 \%$. 

The enhancement of black hole formation at the QCD transition is discussed in
the last section. For standard models of structure formation  without a large
tilt, the amplitudes are not big enough to produce a cosmologically relevant 
amount of black holes \cite{COSMO}. A tilted spectrum could be fine tuned to
produce black holes at the QCD scale. But this tilted spectrum would 
need a break just below the QCD scale in order not to overproduce smaller
black holes, a second fine-tuning. We conclude that the QCD transition
enhances the probability of black hole  formation, but for an observable amount
of black holes today the preexisting spectrum would have to be fine tuned
around the QCD scale, and the major effect would not be due to the QCD
transition.

\section{Supercooling and Sudden Reheating}

The quark-gluon plasma, photons and leptons are tightly coupled via strong,
electromagnetic and weak interactions, $\Gamma/H \gg 1$, and make up a single
perfect (i.e. dissipationless) radiation fluid at scales $\lambda > 10^{-7}
R_{\rm H}$, see \cite{PRD}. The pressure for perfect fluids with negligible
chemical potential only depends on temperature. In the bag model \cite{bag},
which we will consider for simplicity, the quark-gluon plasma (QGP) is
described by $p^{\rm bag}_{\rm QGP} = p^{\rm ideal}_{\rm QGP}-B$, i.e.
\begin{equation}
\label{pQGP}
   p_{\rm QGP} = {\pi^2\over 90} g^*_{\rm QGP} T^4  - B, 
\end{equation}
where $g^*$ is the effective number of relativistic helicity states, and $B$
is the bag constant. Above the phase transition temperature we include $u$,
$d$ quarks and gluons in the quark-gluon plasma and $\gamma, e, \mu ,\nu's$ in
the photon-lepton fluid. Below the phase transition we model the radiation
fluid as a hadron gas (HG) of massless pions, tightly coupled to the photons
and leptons. 

\begin{figure}
\begin{center}
\includegraphics[width=8cm]{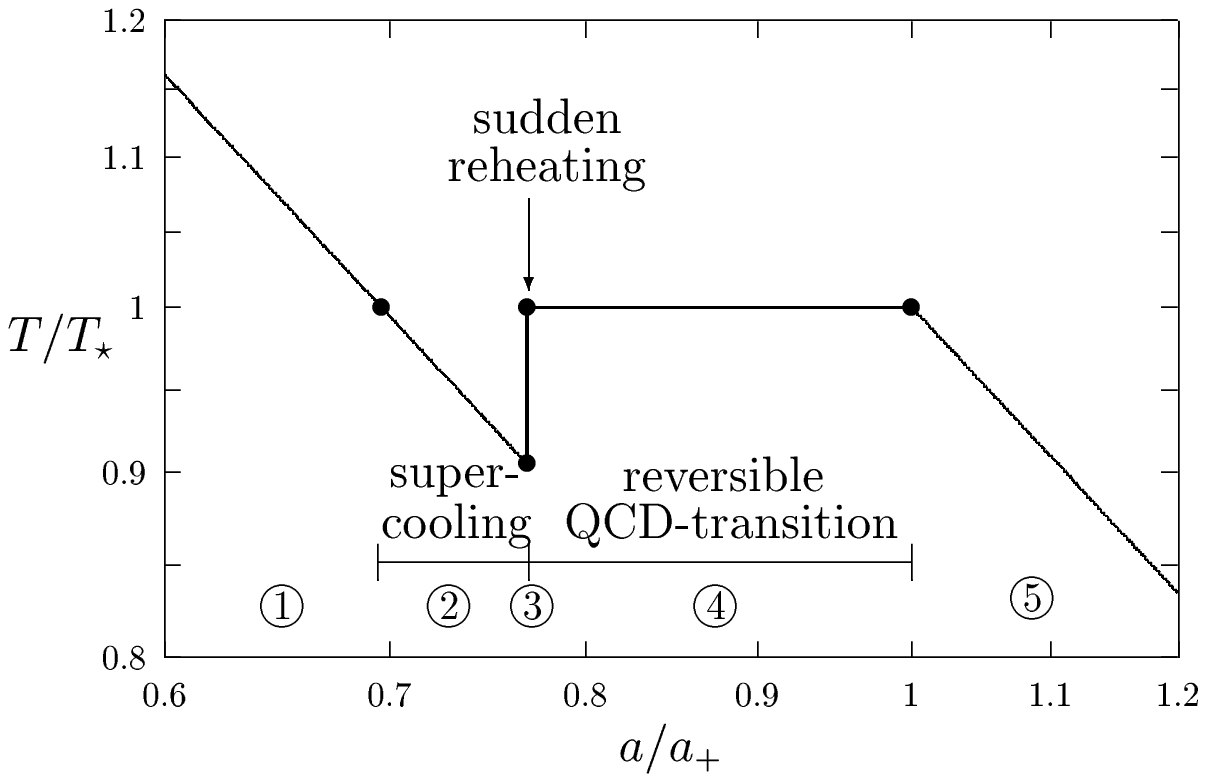}
\end{center}
\caption{\label{fig1}
The evolution of temperature T as a function of the scale factor $a$ in the
bag model. For illustrative purposes the supercooling is $\Delta T/T_\star =10^{-1}$.}  
\end{figure}

In the cosmological expansion the temperature in the bag model drops $\propto
1/a$, see Fig.~\ref{fig1}. To derive this result we use $s = dp/dT$, which
comes from the Maxwell relation for the free energy, hence $s_{\rm QGP}^{\rm
bag}=s_{\rm QGP}^{\rm ideal} \propto T^3$. We further use that the  entropy in
a comoving volume is conserved, $s \propto 1/a^3$, and finally obtain $T
\propto 1/a$.  The energy density is given by $\rho = T s -p$ from the first
law and homogeneity, hence $\rho_{\rm QGP}^{\rm bag} = \rho_{\rm QGP}^{\rm
ideal} + B$. It follows that the sound velocity $(c_s^2)_{\rm QGP} =
(\partial p/ \partial T)_s^{\rm QGP}=1/3$ and that $p/\rho <1/3$.
The bag constant is determined by the transition temperature $T_\star$ and
$\Delta g^* \equiv g^*_{\rm QGP} - g^*_{\rm HG}$ via $p_{\rm
QGP}(T_\star)=p_{\rm HG}(T_\star)$, $B=(\pi^2/90) \Delta g^* T_{\star}^4$.
Similarly the latent heat is given via  $l \equiv T_\star \Delta s$, $l = (2
\pi^2/45)\Delta g^* T_{\star}^4$.

For supercooling, $T<T_\star$, the free energy density $f = -p(T)$ for the QGP
is higher than for the HG, and therefore the QGP is not the equilibrium
state. However the fluid supercools in the metastable quark-gluon phase until
the formation and growth of hadronic bubbles starts reheating the universe at
$T_{\rm sc} < T_\star$. We give a short overview of bubble
nucleation and sudden reheating, a more detailed discussion of bubble
nucleation can be found e.g. in \cite{PRD}. Assuming
homogeneous bubble nucleation (no 'dirt') the probability to
form a bubble is proportional to $\exp(\Delta S)$, where $\Delta S$ is the
change in entropy by creating a bubble. $\Delta S$ is determined by the work
required to form a (spherical) bubble, $-T_{\star} \Delta S = \Delta F
=(p_{\rm QGP} - p_{\rm HG}){4\pi\over 3} R^3 + \sigma 4\pi R^2$,
where $\sigma$ is the surface tension of a hadronic bubble. Only for
large enough bubbles ($R \ge R_{\rm crit}$) free energy is gained. The 
probability to form a hadronic bubble with critical radius per unit volume and unit time is given by
\begin{equation} 
\label{I}
I(T) = I_0 \exp\left( - {\Delta F_{\rm crit}\over T}\right) \ ,
\end{equation}
with $\Delta F_{\rm crit} = 16\pi\sigma^3/[3(p_{\rm HG} - p_{\rm QGP})^2]$. 
For small supercooling, $\eta \equiv 1 -  T/T_\star \ll 1$, we obtain $
(p_{\rm QGP} - p_{\rm HG}) \approx (T-T_\star) \Delta (dp/dT) = - \eta T_\star
\Delta s = - \eta l $, where $l$ is the latent heat.
Therefore the probability to form a critical bubble per unit volume and
unit time can be written as 
\begin{equation}
I \approx T_\star^4 \exp\left(-A/\eta^2\right),
\end{equation}
with $A \equiv 16 \pi \sigma^3/(3 l^2 T_\star)$. A numerical prefactor to
$T_\star^4$ would be irrelevant for our purposes.

 The surface tension $\sigma$ and the latent heat $l$ are the crucial
parameters  for $T_{\rm sc}$. Data are available only for quenched lattice QCD
(gluons only, no quarks) \cite{Iwasaki}: $l/T_\star^4 \approx 1.40 (9)$ and a
very small surface tension $\sigma/T_\star^3 \approx 0.0155 (16)$. There are
no values for unquenched QCD available yet. Using the results from quenched
lattice QCD we find $A = 2.9 \times 10^{-5}$. 

The bubbles grow most probably by weak deflagration
\cite{Kurki-Suonio,Kajantie,Ignatius}, and the latent heat released
from  the bubbles is distributed into the surrounding QGP by acoustic waves and
by neutrinos, whose mean free path is $10^{-7} R_{\rm H}$, of the order of the
bubble separation. This reheats the QGP to $T_\star$ and  
prohibits further bubble formation.  
 The supercooling temperature fraction $\eta$ can be estimated by the
schematic case of one single bubble nucleated per Hubble volume and per Hubble
time, 
\begin{equation}
\label{eta_sc}
  \eta \approx \left[\frac{A}{4 \ln(T_\star/H_\star)} \right]^{1/2}
\approx 4 \times 10^{-4}  \vspace{2cm}.
\end{equation} 
This rough estimate gives a final supercooling fraction which is only
$\approx 20 \%$ too low compared to the more realistic case of one bubble
nucleated per ${\rm cm}^3$ per $10^{-6}$ of a Hubble time, see \cite{PRD}. The
time needed for the supercooling is given by $\Delta t_{\rm sc}/t_{\rm H} =
\eta/(3 c_s^2)= {\cal O}(10^{-3})$. In the figures we took $\eta = 10^{-1}$
for illustrative purposes. At the maximal supercooling the universe suddenly
reheats up to the QCD transition temperature $T_\star$. The sudden reheating
implies a jump in pressure $\left[p\right] \equiv p_{\star} - p_{\rm sc}$.  
\begin{equation}
  {\left[p\right] \over\rho+p} = \eta,
\end{equation}
up to linear order in $\eta$. This jump in pressure depends on the
maximal supercooling and can be a huge effect. In the case of the cosmological
QCD transition it is $\approx 100$ times larger than preexisting COBE
normalized perturbations. On the other hand, the jump in entropy density is
small since it has to be quadratic in $\eta$ due to the second law,
$\left[s\right]/s = \frac32 \eta^2$.  

The reheating is treated as a sudden process happening at the maximal
supercooling, see Fig.~\ref{fig1}. This is justified since the reheating time
is $\approx 10^{-6} t_H \ll \Delta t_{\rm sc}$, which follows from the following
consideration. Bubbles present at a given time have been nucleated typically
during the preceding time interval $\Delta t_{\rm nucl} \equiv  I/({\rm
d}I/{\rm d} t)$. Using the relation between time and supercooling 
$\eta$, ${\rm d}\eta/{\rm d}t = 3 c_s^2 /t_{\rm H}$, we find 
\begin{equation}
  \Delta t_{\rm nucl}/t_{\rm H}= \eta^3/(6 A c_s^2)={\mathcal O}(10^{-6}).
\end{equation}
The corresponding bubble nucleation distance is a few cm \cite{Fuller,Christiansen}. 

During the reversible part of the first order phase transition thermal
equilibrium between the quark-gluon plasma and the hadron gas is maintained.
We showed in \cite{PRL} that pressure gradients and the isentropic sound
speed (for wavelengths $\lambda$ much larger than the bubble separation),
$c_s=\left(\partial p/\partial \rho \right)_S^{1/2}$, must be zero during a
first-order phase transition of a fluid with negligible chemical potential
(i.e. no relevant conserved quantum number). The sound speed must be zero,
because for such a fluid the pressure can only depend on the temperature,
$p(T)$, and because the transition temperature $T_{\star}$ has a given value,
it cannot depend on any parameter, hence $p(T_{\star}) = p_{\star}$ is a given
constant, and  $c_s=0$. During the entire QCD transition the sound speed stays
zero and suddenly rises back to the radiation value $c_s=1/\sqrt{3}$ after the
transition is completed. In contrast pressure varies continuously and goes
below the ideal radiation fluid value $p=\rho/3$, but stays positive. 

\section{Amplification of Subhorizon Perturbations}

The evolution of linear cosmological perturbations is first analyzed on
subhorizon scales, where Newtonian concepts for space and time (for an
expanding radiation fluid) are applicable. The perturbations on subhorizon
scales get an amplification factor which is quadratic in wave number
$k_{\rm}$ for $k \gg H$.   

\begin{figure}
\begin{center}
\includegraphics[width=8cm]{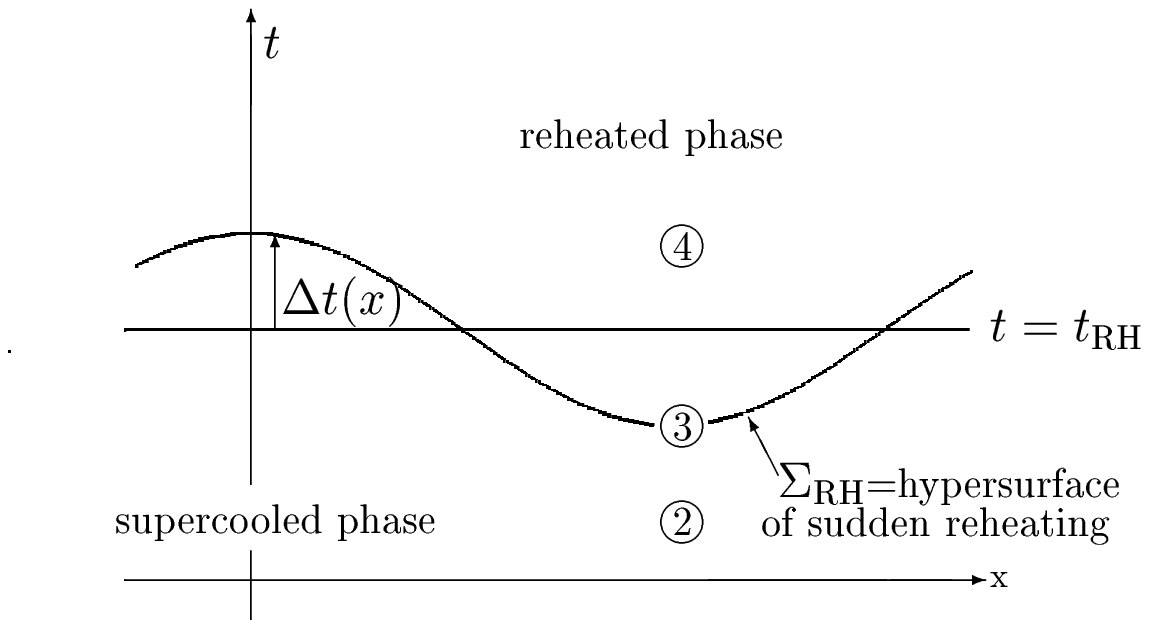}
\end{center}
\caption{\label{fig2} 
A space-time diagram at the time of sudden reheating: The
supercooled (2) and the reheated (4) epoch are separated by the hypersurface
of reheating $\Sigma_{\rm RH}$ (3). On $\Sigma_{\rm RH}$ temperature, hence
pressure, jumps uniformly. $\Delta t({\bf x})$ denotes the lapse of time
compared to the average time, $t_{\rm RH}$, for a certain fluid element to
reach the hypersurface of reheating. For subhorizon wavelengths $t_{\rm RH}$
is the average Newtonian time for reheating, for horizon and superhorizon
wavelengths (general relativistic case),  $t=t_{\rm RH}$ is a
hypersurface of constant time in a given gauge, e.g. in the uniform expansion
gauge.}   
\end{figure}

The maximal supercooling is reached when bubble
nucleation becomes efficient and the radiation fluid suddenly reheats, i.e. it
happens on a hypersurface of constant temperature, $T=T_{\rm sc}$, hence
uniform energy density. On this hypersurface of locally sudden
reheating, $\Sigma_{\rm RH}$, pressure jumps uniformly, see Fig.~\ref{fig2}.

At a given Newtonian time, overdensities have higher temperature than
underdensities and therefore reheat later. This time delay (lapse of time) of
the actual reheating of a fluid element on $\Sigma_{\rm RH}$ from the average
Newtonian time of reheating, $t_{\rm RH}$, is denoted by $\Delta t({\bf x})$
and shown in Fig.~\ref{fig2} for an overdensity of one perturbation mode
$k$. The time delay $\Delta t({\bf x})$ follows from the condition of uniform
energy density on the surface of reheating, $\rho({\bf x},t+\Delta t({\bf
x}))=$ uniform,  
\begin{equation}
  \left. \epsilon({\bf x})\right|^{(-)} + \Delta t({\bf x}) 
   \left.{d \rho \over d t} \right|^{(-)} = 0.
\end{equation}
$\epsilon({\bf x}) \equiv \delta \rho$ denotes the energy density perturbation
at a given time, $(-)$ means fixed time immediately before the short period
during which the various fluid elements reheat. Inserting the
continuity equation for the FRW background $d \rho / d t= - 3 H (\rho +
p)$ at the time of maximal supercooling gives 
\begin{equation}
\label{timelapse}
   \left. \Delta t({\bf x}) = \frac{1}{3H} 
             {\epsilon({\bf x}) \over \rho + p}
             \right|^{(-)}.
\end{equation} 
One can apply the same argument for the perturbations and the
background just after reheating. The time delay of $\Sigma_{\rm RH}$ is the
same whether evaluated before or after the hypersurface, and this condition
gives the discontinuity equation for $\epsilon$, 
\begin{eqnarray}
  \label{jump}
 \left[{\epsilon \over \rho + p}\right] \equiv \left.{\epsilon \over \rho +
p}\right|^{(+)} - \left.{\epsilon \over \rho + p}\right|^{(-)} = 0.
\end{eqnarray} 
$\rho$ is continuous at the actual hypersurface
$\Sigma_{\rm RH}$, but the effects of the fluid expansion is different in the
two media. This generates an effective discontinuity in $\epsilon$.

We now integrate the evolution equations for the relativistic fluid over the
very short time period of order $\Delta t$ when reheating happens at some
${\bf x}$. We split $\rho$ and $p$ into a homogeneous term and an
inhomogeneous term, $\rho({\bf x},t) = \rho(t) + \epsilon({\bf x},t)$, $p({\bf x},t) = p(t) + \pi({\bf x},t)$. In distinction to usual
perturbation theory the pressure inhomogeneities compared to the background
energy density $\rho$ are of order $\Delta T/T_\star$. The pressure
inhomogeneities are the dominant terms, the driving terms, all remaining terms
can be neglected during the short time interval when reheating occurs at
various ${\bf x}$. In this approximation the continuity equation (energy
conservation) and the Euler equation (momentum equation) read, 
\begin{mathletters}
\begin{eqnarray}
  \label{Csub}
  \partial_t \epsilon({\bf x},t) &=& - 3H \pi({\bf x},t)\\
  \label{Esub}
  \partial_t {\bf S}({\bf x},t)  &=& - {\bf \nabla} \pi({\bf x},t),
\end{eqnarray} 
\end{mathletters}
where ${\bf S}$ is the momentum density and ${\bf S} = (\rho + p) {\bf v}$ for
a radiation fluid.

The pressure inhomogeneity $\pi({\bf x},t)$ is a sequence of step functions
with step size $\left[p\right]$, while ${\bf \nabla} \pi({\bf x},t)$ is a
sequence of Dirac delta functions. It is necessary to distinguish the
unperturbed pressure $p(t)$, which makes a jump $\left[p\right]$ at $t=t_{\rm
RH}$, from the average pressure $\bar{p}(t)$, which does not make a jump at
$t=t_{\rm RH}$. It is very convenient to define the pressure perturbations as
$\pi({\bf x},t) \equiv p({\bf x},t) -p(t)$, therefore $\pi({\bf x},t)  \ne
p({\bf x},t)-\bar{p}(t)$. The dominant pressure inhomogeneity is 
\begin{equation}
\label{pofx}
      \pi({\bf x},t)=  \left[p\right]
      \{\theta(t-t_{\rm RH}-\Delta t({\bf x}))-
                   \theta(t-t_{\rm RH})\},
\end{equation}
with $\left[p\right]\equiv p^{(+)}-p^{(-)}$.  
Integrating the continuity equation Eq.~(\ref{Csub}) over the short period of
huge pressure differences gives
the jump condition for $\epsilon$,
\begin{equation}
\label{jumpE}
\left[\epsilon({\bf x})\right] =  \left[p\right] 3 H \Delta t({\bf x})
=\left.\left[p\right] {\epsilon({\bf x}) \over \rho + p} \right|^{(-)},
\end{equation}
in agreement with Eq.~(\ref{jump}). 

Integrating the Euler equation Eq.~(\ref{Esub}) over the short period of huge
pressure differences, gives the jump condition for the momentum density ${\bf S}$,
\begin{mathletters}
\begin{eqnarray}
\label{jumpS1}
\left[{\bf S}({\bf x}) \right]&=&\left[p\right] \int dt 
            \delta(t-t_{\rm RH}-\Delta t({\bf x})){\bf \nabla} (\Delta t) \\
\label{jumpS}
            &=&\left[p\right] \frac{1}{3 H} {\bf \nabla} 
            \left.{\epsilon({\bf x}) \over \rho + p} \right|^{(-)}. 
\end{eqnarray}
\end{mathletters}
The discontinuity in ${\bf S}$ is a local discontinuity at the actual
hypersurface of reheating (produced by the $\delta$ functions of
Eq.~(\ref{Esub}) and Eq.~(\ref{jumpS1})). It is a sudden impulse imparted on
each fluid element.

In the evolution equations Eqs.(\ref{Csub}), (\ref{Esub}) different Fourier
modes are coupled during the short period of reheating. Therefore the standard
procedure to evolve each ${\bf k}$ mode separately is not applicable at
reheating. However the resulting jump conditions obtained by working in ${\bf
x}$ space, Eqs.(\ref{jumpE}), (\ref{jumpS}), are valid for each ${\bf k}$ 
mode separately.   

On subhorizon scales the discontinuity in $\epsilon$ is independent of $k$
(for a preexisting Harrison-Zel'dovich spectrum), the discontinuity in ${\bf
S}$ is proportional to $k_{\rm phys}$ and is the dominant effect in the
subhorizon limit. 

\begin{figure} 
\begin{center}
\includegraphics[width=8cm]{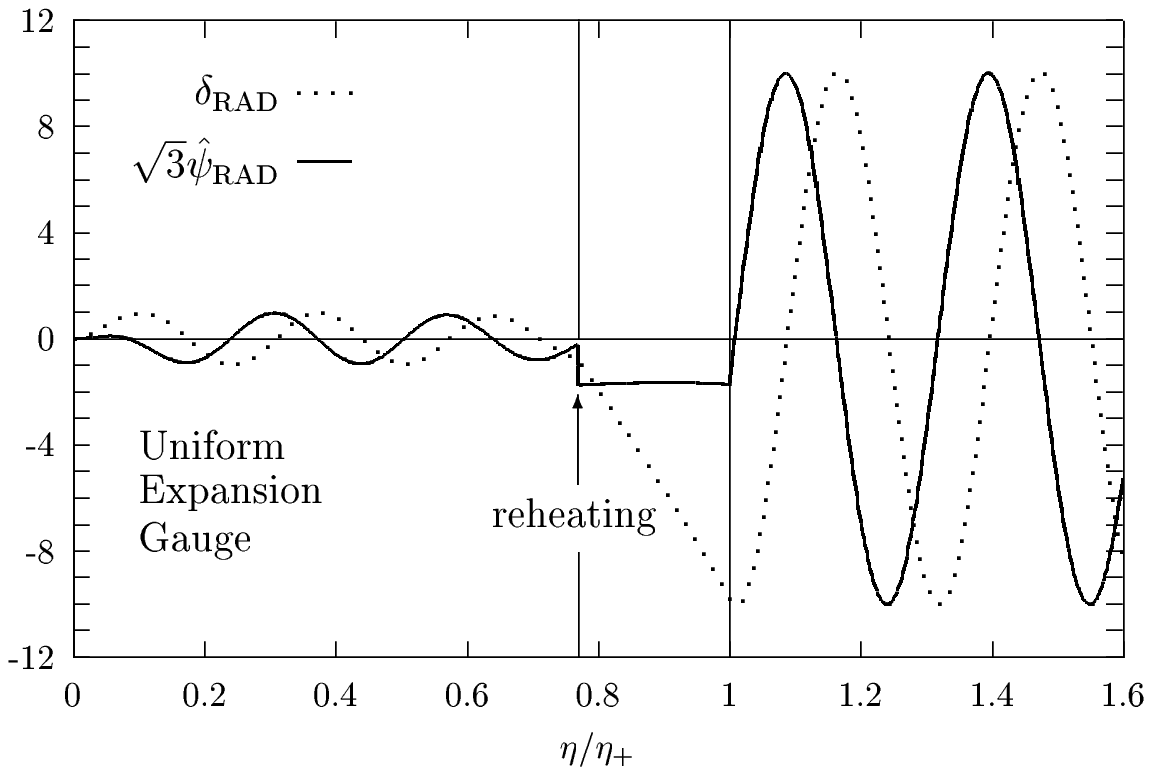}
\end{center}
\caption{\label{fig3}
The evolution in conformal time $\eta$ of the density contrast ($\delta_{\rm
RAD}\equiv \epsilon/\rho$) and the velocity ($\hat{\psi}_{\rm RAD}\equiv $
$i {\bf k} \cdot {\bf S}/(k \rho)=(\rho+p) v/\rho$) for the radiation fluid
in uniform expansion (Hubble) gauge. At sudden reheating, the fluid velocity
jumps. During the reversible part of the QCD transition, marked by the 2
vertical lines, the velocity stays approximately constant and the density
contrast grows linearly. The incoming amplitude is normalized to $1$.
}
\end{figure}

The evolution of density perturbations before and after sudden reheating
can be computed in a Jeans analysis generalized to a relativistic fluid for
each Fourier mode separately, see \cite{PRL}. The preexisting subhorizon
perturbations before sudden reheating are acoustic oscillations, see
Fig.~\ref{fig3}. In the bag model the sound speed $c_s=1/\sqrt{3}$ and the
amplitudes $A_{\rm in}$ for $\epsilon$ and $|\sqrt{3} {\bf S}|$ are equal
until sudden reheating. At $t_{\rm RH}$, $\epsilon$ and ${\bf S}$ jump
according to Eqs.(\ref{jumpE}), (\ref{jumpS}). After sudden reheating, during
the reversible part of the QCD transition, the sound speed $c_s \equiv 0$,
i.e. the restoring force in the acoustic oscillations vanishes as first shown
in \cite{Helv,PRL}, and density perturbations go into free fall. Since the QCD
transition lasts less than a Hubble time, gravity is negligible for subhorizon
scales and density perturbations have approximately constant velocity,
$\epsilon(t) = \epsilon^{(+)} - (t-t_{\rm RH}){\bf \nabla}
{\bf S}^{(+)}$. This  gives an amplification which is linear in
$k_{\rm phys}$ and proportional to ${\bf S}^{(+)}$. After the
QCD transition is completed, one has acoustic oscillations again and the
total amplification factor for $k \gtrsim H$ is
\begin{equation}
\label{k1}
 {A_{\rm out}\over A_{\rm in}} 
  = {k \over k_1}|{k \over k_2} \cos \varphi_{\rm in} - \sin
\varphi_{\rm in}| \ .
\end{equation}
$\varphi_{\rm in}$ is the phase of the incoming acoustic oscillation at $t_{\rm
RH}$. The scale $k_1 \equiv \sqrt{3}/ \Delta t_{\rm trans}\approx R_H^{-1}$
depends on the duration of the reversible QCD transition, $\Delta t_{\rm
trans}$. This contribution is due to the free fall and starts just below the
Hubble scale, see Fig.~\ref{fig4}. $k_2$ depends on the amount of
supercooling, $k_2 \equiv H_{\rm RH}2\sqrt{3}/\eta$. 
The amplification factor is quadratic in $k$ for $(\lambda/R_H)<\eta$. 
 
Jump conditions like Eqs.(\ref{jumpE}), (\ref{jumpS}) must be used whenever
there is a jump in the background quantities (as opposed to a jump only in
$c_s^2$ as in the case without supercooling and sudden reheating). Only the
background pressure is allowed to jump not the energy density, according to the
continuity equation. Due to the second law of thermodynamics a fluid can only
heat instantaneously, sudden cooling is impossible.  

\section{General Relativistic Analysis}

A slicing of space-time (with space-like hypersurfaces on which time is
defined to be constant) is needed to formulate time-evolution equations in
general relativity. We restrict ourselves to time-orthogonal foliations. For
linear cosmological perturbations a choice of a foliation already fixes a
gauge. For sudden reheating there is one special slice, the hypersurface of
reheating $\Sigma_{\rm RH}$, which is a surface of constant energy density. It
is a fixed-time slice in uniform density (UD) gauge, see \cite{Bardeen}. 

We use the uniform density gauge, the gauge adapted to our problem, to evolve
through sudden reheating. We now show that in UD gauge the relevant variables
are continuous (do not jump). The hypersurface of sudden reheating,
$\Sigma_{\rm RH}$, has definite extrinsic curvature $K_{ij}$ and intrinsic
curvature $^{(3)}R_{ij}$ at each point on $\Sigma_{\rm RH}$, hence these
quantities are continuous at reheating. The momentum density ${\bf S}$ in UD
gauge also stays continuous at reheating due to the momentum constraint of
general relativity, $D_j K^j_{\;\;i}-D_i K^j_{\;\;j}=8 \pi G S_i$, see
e.g. \cite{Wald}. The energy density is uniform in space and continuous in time
in UD gauge.

The UD gauge is not suitable for evolving before and after reheating, because
the UD gauge is a singular gauge in the subhorizon limit, $H/k_{\rm phys}
\rightarrow 0$, see below. Therefore we shall solve the dynamics before and
after sudden reheating using the uniform expansion gauge (UE), which is
nonsingular both in the superhorizon and subhorizon limits and in which the
physical dynamics will turn out to be most transparent (in contrast to the UD
gauge). However in the UE gauge the relevant variables are discontinuous
(jump) at sudden reheating. The uniform (Hubble) expansion gauge is defined by
requiring that the fundamental observers, who are at rest on the slice,
$\underline{u}({\rm obs}) = \underline{n}(\Sigma)$, have uniform Hubble
expansion, i.e. the perturbation of the mean extrinsic curvature of $\Sigma$,
$\delta \left[{\rm tr} K^i_{\:j}(\Sigma)\right] \equiv \kappa$, is zero,
$\kappa_{\rm UE} \equiv 0$.

The geometrical properties of a hypersurface $\Sigma$ in the longitudinal
sector are given by the following variables, Bardeen 1989 \cite{Bardeen}: The
perturbation of the trace of the extrinsic curvature 
$\delta \left[ {\rm tr}K^i_{\;j}(\Sigma)\right] \equiv \kappa$, the traceless
part of the extrinsic curvature (shear $\sigma^i_{\rm j}$ of normals),
which is determined by $\chi$ through $\sigma^i_{\;
j}(\Sigma)=-\left(\partial^i\partial_j-\frac13 \delta^i_j \bigtriangleup
\right)\chi$, and $^{(3)}R(\Sigma) = - 4 \bigtriangleup \varphi$. The lapse
of time between the perturbed hypersurfaces is $\left(1+\alpha\right)$. The
connection between the geometric variables and the gauge potentials (in the
longitudinal sector with time-orthogonal coordinates),
\begin{equation}
 ds^2 = -\left( 1+ 2 \alpha \right) d t^2 + a(t)^2 \left[ \delta_{i j}(1+2
\varphi) +2 \partial_i\partial_j \gamma \right] dx^i dx^j,
\end{equation}
is as follows: $\chi = a^2 \dot{\gamma}$, $\kappa = -3(\dot{\varphi}-H\alpha)
- \bigtriangleup \chi$.

The state of the fluid (before or after sudden reheating) on a hypersurface of
fixed time (in some gauge) is given by the following variables: The energy
density perturbation $\delta \rho \equiv \epsilon$ and the momentum density
${\bf S}$, which can be written as ${\bf S}= {\bf \nabla} \psi$ in the
longitudinal sector (sector of scalar perturbations). $p(\rho)$ is given by
the bag model before and after the irreversible reheating. There are no
anisotropic stresses for our perfect fluid (QCD, $\gamma$, leptons). The
total set of variables in the longitudinal sector is
$\left(\epsilon,\psi\right)$, $\left(\kappa,\chi,\varphi\right)$, $\alpha$. 

For the dynamical equations in UE gauge we choose to take the evolution
equations of the fluid, $\nabla_\mu T^{\mu \nu} = 0$, which are the continuity
equation and (in the longitudinal sector) the $3$-divergence of the Euler
equation,
\noindent
\begin{mathletters}
\begin{eqnarray}
\label{CUEG}
\partial_t \epsilon &=& - 3 H(\epsilon + \pi) - \bigtriangleup \psi -
3 H (\rho + p)\alpha \\
\label{EUEG}
\partial_t \psi &=& - 3 H \psi - \pi - (\rho + p)\alpha,
\end{eqnarray}
with $\pi \equiv \delta p$. The system of dynamical equations is closed by
Einstein's $R_{\hat{0}\hat{0}}$-equation,
\begin{eqnarray}
\label{PUEG}
(\bigtriangleup + 3 \dot{H})\alpha = 4\pi G (\epsilon + 3 \pi), 
\end{eqnarray} 
\end{mathletters}
together with the equation of state. This set of equations has exactly the
same structure as the Jeans equations: Two first-order time-evolution
equations for the fluid state variables $(\epsilon, \psi)_{\rm UE}$, whose
initial values can be chosen free of constraints, supplemented by an elliptic
equation for $\alpha_{\rm UE}$, which plays the role of the gravitational
potential. The lapse $\alpha$ is the only geometrical variable needed to solve
the dynamical evolution in UE gauge. The other geometric perturbations,
i.e. the shear $\chi$ and the intrinsic curvature $\varphi$, do not appear in
our closed set of evolution equations, Eqs.~(\ref{CUEG}-\ref{PUEG}). If
desired, $\varphi$ and $\chi$ can be computed from the state variables via the
energy and momentum constraints, $\bigtriangleup \varphi = - 4 \pi G \epsilon$,
$\bigtriangleup \chi = - 12 \pi G \psi$ on any UE slice.

To make the gauge transformation from the uniform expansion gauge (independent
variables $\epsilon, \psi$) to the uniform density gauge (independent
variables $\psi,\kappa$) we need the gauge transformation formulae for
$\left(\epsilon,\psi,\kappa\right)$ \cite{Bardeen}  
\noindent
\begin{mathletters}
\begin{eqnarray}
\label{EGauge}
\tilde{\epsilon} - \epsilon &=&  - 3 H (\rho + p) \Delta t \\
\label{PGauge}
\tilde{\psi} - \psi &=&  - (\rho + p) \Delta t\\
\label{KGauge}
\tilde{\kappa} - \kappa&=& - (3\dot{H} - k_{\rm phys}^2)  \Delta t \ ,
\end{eqnarray}
\end{mathletters}
where the variables with tilde correspond to the new gauge and $\Delta t$ is
the lapse of time from the old to the new hypersurface, see
Fig.~\ref{fig2}. 

We now derive the discontinuity conditions at sudden reheating for the uniform
expansion gauge. We use the state variables
$\left(\epsilon,\psi\right)^{(-)}_{\rm UE}$ just before reheating and use
Eq.~(\ref{EGauge}) with $\tilde{\epsilon}_{\rm UD}\equiv 0$ to obtain
\noindent
\begin{eqnarray}
\label{timedelay}
\Delta t^{(-)} = {\epsilon_{\rm UE} ^{(-)} \over 3 H (\rho + p^{(-)})},
\end{eqnarray}
where $(-)$ means $t_{\rm UE}$ immediately before reheating. The momentum
potential $\psi_{\rm UD}$ and the extrinsic curvature $\kappa_{\rm UD}$ follow
from Eqs.(\ref{PGauge}), (\ref{KGauge}), 
\noindent 
\begin{mathletters}
\begin{eqnarray} 
\label{PUD} 
\psi_{\rm UD} &=& \psi_{\rm UE}^{(-)} - {\epsilon_{\rm UE}^{(-)} \over 3H} \\
\label{KUD} 
\kappa_{\rm UD}&=& - (3\dot{H}^{(-)} - k_{\rm phys}^2) {\epsilon_{\rm 
UE}^{(-)} \over 3H (\rho + p^{(-)})} .  
\end{eqnarray} 
\end{mathletters} 
$\psi_{\rm UD}$ and $\kappa_{\rm UD}$ stay continuous at sudden reheating. We 
make the analogous gauge transformation (from UE to UD) after sudden 
reheating. From the $\kappa_{\rm UD}$ equation we get the discontinuity 
condition for $\epsilon$, 
\noindent 
\begin{eqnarray} 
\label{Ejump} 
\left[{3 \dot{H} - k_{\rm phys}^2 \over \rho+p} \epsilon_{\rm UE} \right]&=&0,
\end{eqnarray} 
where the jump in the background quantity $\dot{H} = -4 \pi G (\rho +p)$ is 
the given input, and the jump in $\epsilon$ is the output. In the subhorizon 
limit $\left[\epsilon/(\rho+p)\right] = 0$ and $\Delta t^{(-)} = \Delta 
t^{(+)}$, i.e. we recover the results of the analysis using 
Newtonian geometry, Eq.~(\ref{jumpE}). In the superhorizon limit $\epsilon$ does not jump, 
$\left[\epsilon \right] =0 $, but  $\Delta t^{(-)} \ne \Delta t^{(+)}$. This 
means that the coordinate regions $t_{\rm UE}<t_{\rm RH}$ and $t_{\rm
UE}>t_{\rm RH}$ cover certain regions of space-time doubly and other regions
not at all. Applying the $\psi_{\rm UD}$ equation before and after reheating
we obtain 
 \noindent
\begin{eqnarray}
\label{Sjump}
\left[\psi_{\rm UE} \right]&=&\frac{1}{3 H}\left[\epsilon_{\rm UE} \right].
\end{eqnarray}
The resulting modification of the
perturbation amplitude due to the QCD phase transition (transfer function) is
shown in Fig.~\ref{fig4}. 

On superhorizon scales $(\epsilon, \psi)_{\rm UE}$ do not jump. An even
stronger statement can be made for $\delta_{\rm UE} \equiv \epsilon_{\rm
UE}/\rho$: The evolution of the growing mode (i.e. growing in relative
importance) in $\delta_{\rm UE}$ is given by 
\begin{equation}
\label{growing}
\left(\delta_{\rm UE}\right)_{\rm growing \; mode} = {\rm const} \left({k_{\rm
phys}\over H}\right)^2
\end{equation}
for scales $k_{\rm phys}^2 \ll \{H^2,|\dot{H}|\}$ and $p/\rho = w <1$. Written
in this form the growing mode of the state variable $\delta_{\rm UE}$ is
manifestly unaffected during phase transitions or any other changes in the
equation of state. The proof goes as follows. We take the $k_{\rm phys}^2 \ll
|\dot{H}|$ limit of the $R_{\hat{0} \hat{0}}$-equation, $-3(\rho+p) \alpha =
(\epsilon + 3 \pi)$, and we insert this into the continuity equation,
Eq.~(\ref{CUEG}), and into the Euler equation, Eq.~(\ref{EUEG}),
\noindent
\begin{mathletters} 
\begin{eqnarray}
\label{Csuper}
\partial_t \epsilon + 2 H \epsilon &=& k_{\rm phys}^2 \psi\\
\label{Esuper}
\partial_t \psi     + 3 H \psi     &=& \frac13 \epsilon.
\end{eqnarray}
\end{mathletters}
We note that the pressure gradients have dropped out. For the mode which will
turn out to be the growing mode the two terms on the left-hand side of the
Euler equation do not cancel, and $\epsilon$ is relevant in the Euler
equation, hence $\psi = {\mathcal O}\left(\epsilon/H\right)$. Therefore the
$\psi$ term is negligible in the continuity equation for the growing mode
if $k_{\rm phys}^2\ll H^2$, $\partial_t (\ln \epsilon) = - 2 \partial_t (\ln
a)$. Hence  $\epsilon = {\rm const \;} a^{-2}$ and using the Friedmann equation
$\delta_{\rm UE} \equiv \epsilon_{\rm UE}/\rho = {\rm const \;} (k_{\rm
phys}/H)^2$. For the decaying mode the two terms on the left-hand side of the
continuity equation do not cancel, and $\psi$ is relevant in the continuity
equation, hence $\epsilon = {\mathcal O}\left(k_{\rm
phys}^2\psi/H\right)$. Therefore the $\epsilon$ term is negligible in the Euler equation for the decaying mode, $\partial_t (\ln \psi) = - 3 \partial_t (\ln
a)$. Hence  $\psi = {\rm const \;} a^{-3}$, and the continuity equation gives
$\epsilon \propto a^{-2} \int a^{-3} dt$. With $a \propto t^{2\over 3(1+w)}$
it follows that this is indeed the decaying mode for $w\equiv p/\rho <1$. The
law, Eq.~(\ref{growing}), can be translated (using the energy constraint) into
the 'conservation law' $\left(\varphi_{\rm UE}\right)_{\rm growing \; mode}={\rm const}$. This is consistent with the 'conservation law' \cite{BST}
$\zeta_{\rm growing \; mode}= {\rm const}$ for $k_{\rm phys}^2
\ll H^2$, where $\zeta \equiv \varphi +{\epsilon \over 3(\rho +p)}$.

Deruelle and Mukhanov \cite{Mukhanov} have analyzed jump conditions (matching
conditions) in zero shear (ZS) gauge, also called longitudinal gauge, where
the shear of the normals to the equal time hypersurfaces is zero by
definition. They focussed on superhorizon physics. The matching conditions are
much more complicated in zero shear gauge than in uniform expansion gauge; no
superhorizon discontinuities in UE state variables, complicated discontinuity 
conditions in ZS in Ref.~\cite{Mukhanov}. This is connected to the fact that
the zero shear gauge is singular in the superhorizon limit (Bardeen 1988
\cite{Bardeen}), and one particular aspect of this is that $\Delta t_{\rm ZS,
UD}/\Delta t_{\rm UE,UD}\propto \left(H/k_{\rm phys}\right)^2 \rightarrow
\infty$ in the superhorizon limit. 

Finally we show that the uniform density gauge is singular in the subhorizon
limit. For a Harrison-Zel'dovich spectrum the subhorizon density contrast
$\delta_{\rm UE}\equiv \epsilon/\rho={\mathcal O}(10^{-4})$ is independent of
$k$, the state variables $(\epsilon,\psi)_{\rm UE}$ agree with
$(\epsilon,\psi)$ in a Jeans analysis. From Eq.~(\ref{timedelay}) and the
subhorizon limit of Eq.~(\ref{KUD}) we obtain $\kappa_{\rm UD}/H={\mathcal
O}(k^2_{\rm phys} \delta_{\rm UE}/H^2)\rightarrow\infty$. Eq.~(\ref{PUD})
gives ${\bf S}_{\rm UD}/\epsilon_{\rm UE}={\mathcal O}({\bf k}_{\rm
phys}/H)\rightarrow\infty$. 

\section{Black Hole Formation at the QCD Transition ?}

\begin{figure} 
\begin{center}
\includegraphics[width=8cm]{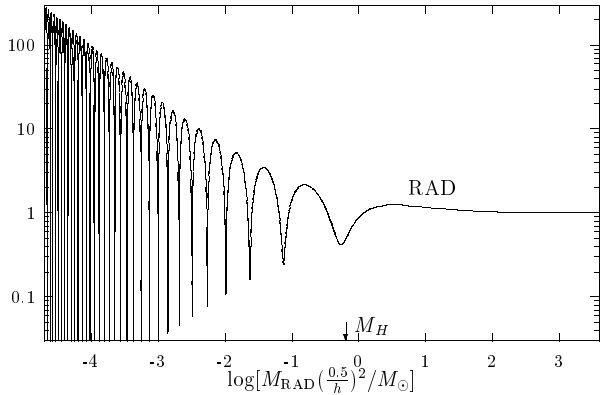}
\end{center}
\caption{\label{fig4}
The modifications of the radiation fluid amplitude $A_{\rm out}/A_{\rm in}$
(transfer function) due to the QCD transition in the bag model with a
supercooling of $\Delta T/T_\star = 10^{-1}$. On the horizontal axis the
wavenumber $k$ is represented by the RAD mass contained in a sphere of radius
$\pi/k$. $M_H$ is the mass inside the horizon at the QCD transition.}
\end{figure}

Black holes form in a radiation dominated universe if the density contrast of
a top hat perturbation inside the Hubble radius is in the range $1/3 \le \delta_{\rm H} \le 1$
\cite{BHF}. For an  observable amount of $1 M_{\odot}$ black holes today,
i.e. $\Omega_{\rm BH}^{(0)}={\cal O} (1)$, the fraction $\beta$ of energy 
density converted to black holes at the QCD transition must be ${\cal
O}(a_{\rm QCD}/a_{\rm equality}) \approx 10^{-8}$. For a gaussian distribution
of density fluctuations the fraction $\beta$ of $\rho _{\rm BH}$ at the time
of formation is given by, 
\begin{equation}
\beta\equiv {\rho_{\rm BH} \over \rho_{\rm tot}} \approx \frac{1}{\sqrt{2 \pi}
\delta_{\rm rms}} \int_{1/3}^1 \exp(- \frac{\delta^2}{2 \delta_{\rm rms}^2}) d
\delta 
\end{equation}
Without any enhancement from the QCD transition this requires $\delta_{\rm
rms} \approx 0.06$ \cite{Bullock}. The sudden reheating at the onset of a
first-order QCD transition leads to huge amplifications of density
perturbations on scales far below the Hubble horizon. But at the horizon scale
the QCD transition gives enhancement factors of $2.0$ and $2.2$ for the bag
model without resp. with sudden reheating for a supercooling of $10 \%$,
Fig.~\ref{fig4}, an additional enhancement of $10\%$. For lattice QCD without
sudden reheating the enhancement factor is $1.5$ \cite{PRL} in our linear
perturbation treatment.
This indicates a corresponding reduction in the required preexisting
perturbation spectrum at the solar mass scale. Cardall and Fuller
\cite{Cardall} used a qualitative argument of Carr and Hawking
\cite{CarrHawking} and the bag model and also obtained a factor $2$ reduction
in the required preexisting perturbation spectrum. These QCD factors of $1.5$
or $\sim 2$ are so modest that a preexisting Harrison-Zel'dovich spectrum
with COBE normalization is very far from giving a cosmologically relevant
amount of black holes \cite{COSMO}. One would have to put in a fine-tuned tilt  $(n-1) \approx 0.36$
to get the desired amount of black holes. However, this tilted spectrum would
overproduce primordial black holes on scales which are only a factor $50$
below the Hubble radius at the transition. Therefore a break in the
preexisting spectrum below the QCD scale would be required, a second fine
tuning.  
 
We conclude that the QCD transition with or without sudden reheating enhances
the probability of black hole formation, but the preexisting spectrum needs to
be fine tuned around the QCD scale, and the major effect would not be due to
the QCD transition.  

\acknowledgements

We thank Dominik Schwarz for useful discussions. P. W. thanks the Swiss
National Science Foundation for financial support.

%%%%%%%%%%%%%%%%%%%%%%%%%%%%%%%%%%%%%%%%%%%%%%%%%%%%%%%%%%%%%%%
%%%%%%%%%%%%%%%%%%%%%%%%%%%%%%%%%%%%%%%%%%%%%%%%%%%%%%%%%%%%%%%

%%%%%%%%%%%%%%%%%%%%%%%%%%%%%%%%%%%%%%%%%%%%%%%%%%%%%%%%%%%%%%%
%%%%%%%%%%%%%%%%%%%%%%%%%%%%%%%%%%%%%%%%%%%%%%%%%%%%%%%%%%%%%%%

%%%%%%%%%%%%%%%%%%%%%%%%%%%%%%%%%%%%%%%%%%%%%%%%%%%%%%%%%%%%%%% 
%%%%%%%%%%%%%%%%%%%%%%%%%%%%%%%%%%%%%%%%%%%%%%%%%%%%%%%%%%%%%%%

%%%%%%%%%%%%%%%%%%%%%%%%%%%%%%%%%%%%%%%%%%%%%%%%%%%%%%%%%%%%%%% 
%%%%%%%%%%%%%%%%%%%%%%%%%%%%%%%%%%%%%%%%%%%%%%%%%%%%%%%%%%%%%%%

\end{document}